\documentclass[prb,aps,showpacs,twocolumn]{revtex4}

\usepackage{epsfig,color}
\usepackage{dcolumn}
\usepackage{amsmath}
\usepackage{color}
\usepackage{amssymb}
\usepackage {graphicx}

\newcommand{\beq}{\begin{equation}}
\newcommand{\eeq}{\end{equation}}
\newcommand{\bea}{\begin{eqnarray}}
\newcommand{\eea}{\end{eqnarray}}
\newcommand{\be}{\begin{equation}}
\newcommand{\ee}{\end{equation}}

\begin{document}

\title{Effect of nodes, ellipticity and impurities on the spin resonance in Iron-based superconductors}
\author {S. Maiti$^{1}$,J. Knolle$^{2}$,I. Eremin$^{3}$ and A.V.~Chubukov$^{1}$}

\affiliation{$^1$ Department of Physics, University of
Wisconsin-Madison, Madison, Wisconsin 53706, USA\\
$^2$Max-Planck-Institut f\"{u}r Physik komplexer Systeme, D-01187
Dresden, Germany\\
$^3$Institut f\"ur Theoretische Physik III,
Ruhr-Universit\"at Bochum, D-44801 Bochum, Germany}

\date{\today}

\begin{abstract}
We analyze doping dependence of the spin resonance of an $s^{\pm}$
superconductor and its sensitivity to the ellipticity of electron
pockets, to magnetic and non-magnetic impurities, and to the angle
dependence of the superconducting gap along electron Fermi
surfaces. We show that the maximum intensity of the resonance
shifts from commensurate to  incommensurate momentum above some
critical doping which decreases with increasing ellipticity. Angle
dependence of the gap and particularly the presence of accidental
nodes lowers  the overall intensity of the resonance peak and
shifts its position towards the onset of the particle-hole
continuum. Still, however, the resonance remains a true
$\delta-$function in the clean limit. When non-magnetic or
magnetic impurities are present, the resonance broadens  and its
position shifts. The shift depends on the type of impurities and
on the ratio of intraband and interband scattering components. The
ratio $\Omega_{res}$/T$_c$ increases almost linearly with the
strength of the interband impurity scattering, in agreement with
the experimental data. We also compare spin response of $s^{\pm}$
and $s^{++}$ superconductors. We show that there is no resonance
for $s^{++}$ gap, even  when there is a finite mismatch between
electron and hole Fermi surfaces shifted by the antiferromagnetic
momentum.
\end{abstract}


\maketitle

\section{Introduction}
\label{sec:1}

The relation between unconventional superconductivity and
magnetism is one of the most interesting topics in
condensed-matter physics. In conventional phonon- mediated
$s-$wave superconductors (SCs), the SC gap $\Delta$ is
approximately a constant along the Fermi surface (FS), and
paramagnetic spin excitations at $T << T_c$ are suppressed below
$2\Delta$ due to formation of Cooper pairs with the total spin
$S=0$. In unconventional SCs, such as layered cuprates or some
heavy fermion materials, the pairing symmetry is $d-$wave, and the
SC gap changes sign along the Fermi surface. In this
situation, the suppression of the spin response is only one effect of superconductivity,
another is the appearance of the resonance below the
onset of the particle-hole continuum, at a set of momenta
which connect FS points with different signs of the
gap.~\cite{eschrig}.  The intensity of the resonance is the
strongest at ${\bf Q}$ for which the gap magnitude is the largest
at the corresponding FS points. For many unconventional SCs such
${\bf Q}$ coincides with the antiferromagnetic momentum
$(\pi,\pi)$.

In general, the spin resonance has contributions from both fermionic excitations in
particle-hole and
particle-particle channels, which are mixed in a
superconductor.~\cite{oleg} In most cases, however, the dominant
contribution to the resonance comes from particle-hole channel and
from this perspective the resonance can be viewed as a spin
exciton~\cite{hao}. The spin resonance is a true
$\delta-$function in a 2D superconductor at $T=0$, but acquires a
finite width at a finite $T$ in 2D, and even at $T=0$ in 3D
systems~\cite{gorkov}, if a line of nodes intersects the locus of FS points
separated by ${\bf Q}$.  A broadened resonance survives even when
the system loses superconducting phase coherence~\cite{millis},
{\it i.e.}, it still exists in some $T$ range above $T_c$, however
its intensity sharply increases only below $T_c$.

Because spin resonance only develops when there is a sign change
of the superconducting gap between ${\bf k}_F$ and ${\bf k}_F +
{\bf Q}$,  it is widely regarded as a probe of unconventional gap
symmetry or structure,  complementary to phase sensitive measurements.
 The observation of the spin resonance
in the cuprates~\cite{rossat} and in some heavy-fermion materials
~\cite{sato,broholm} is a strong evidence of a $d-$wave gap
symmetry in these materials.

The subject of this paper is the analysis of the several features
of the spin resonance seen by inelastic neutron scattering (INS)
in Fe-based superconductors (FeSCs). These systems are
multi-orbital/multi-band metals with two or three hole FS pockets
centered at the $\Gamma$ point $(0,0)$ and two elliptical electron
FS pockets centered at $(\pi,\pi)$ in the actual, folded Brillouin
zone (BZ) with 2 $Fe-$ atoms per unit cell. There also exist FeSCs
with only hole or only electron FSs, but we will not discuss these
materials. Throughout  this paper we assume that FeSCs can be
reasonably well approximated by 2D models and neglect the
anisotropy of electron dispersion along $k_z$. We also consider
only interactions that conserve momentum in the unfolded BZ (1
$Fe-$ atoms per unit cell) and neglect additional interactions via
a pnictogen, which only conserve momentum in the folded BZ and
hybridize the two electron pockets.  This hybridization is
relevant to systems with only electron FSs~\cite{mazin_last}, but
does not look to be important for our consideration.

Angle Resolved Photoemission Spectroscopy (ARPES) studies of FeSCs
with hole and electron pockets demonstrated quite convincingly
that the SC gap in these systems does not have nodes along the
hole FSs, at least for $k_z$ probed by ARPES. The absence of the nodes
is the proof that FeSCs are $s-$wave superconductors, otherwise the gap would
have nodes on hole pockets for all $k_z$. Still, however, in a multi-band SC,
$s-$wave gap can be unconventional and give rise to the spin resonance in INS.
Indeed, s-wave symmetry only implies that the SC gap is
approximately a constant along hole FSs, but does not impose a
restriction on the relative signs of the SC gaps along hole and
electron FSs separated by ${\bf Q}$.  A gap
structure consistent with ARPES can be either a conventional,
sign-preserving  $s^{++}$ gap, or an unconventional $s^{\pm}$ gap
which changes sign between hole and electron FSs.

Both gaps have been proposed for FeSCs based on the two different
assumptions about the interplay between intra-orbital and
inter-orbital screened Coulomb  interactions.  $s^{\pm}$ gap has
been proposed based on the assumption that intra-orbital
interaction is stronger than the inter-orbital one. In the band
description, this condition implies that both intra-pocket and
inter-pocket interactions are positive, i.e., repulsive. Then a conventional
$s-$wave superconductivity is impossible, but  $s^{\pm}$
pairing is possible if  the inter-pocket
is larger than the intra-pocket one. A positive inter-pocket interaction is enhanced by
antiferromagnetic fluctuations and close enough to a magnetic instability
 overcomes intra-pocket repulsion\cite{chubukov,mazin,scalapino,tesanovic,wang,thomale,kuroki,lp}.
The same antiferromagnetic fluctuations also enhance a d-wave
pairing component, but a d-wave gap has  nodes on the hole FSs and
 has a smaller condensation
energy, at least near a magnetic transition.

The alternative, $s^{++}$ pairing has been proposed~\cite{onari} based on the
opposite assumption that  intra-orbital  interaction is weaker
than the inter-orbital one. In the band description, this implies
that intra-pocket interaction is repulsive (positive), but
inter-pocket interactions is attractive (negative). Then  a
conventional $s^{++}$ superconductivity occurs once inter-pocket
interaction exceeds the intra-pocket one. A negative
inter-pocket pairing interaction is enhanced by charge (orbital)
fluctuations in combination with phonons, that is,  if orbital fluctuations play the dominant role,
 the system develops an  $s^{++}$
superconductivity.

Because $s^{\pm}$ gap changes the sign between hole and electron
FSs, it satisfies the same condition for the resonance, $\Delta_{\bf k} =
-\Delta_{\bf k+Q}$ as  in a $d-$wave
superconductor. Accordingly, the spin response of an $s^{\pm}$
superconductor below $T_c$  should contain a sharp, nearly
$\delta-$functional spin resonance~\cite{korshunov,maier,hu}. No
such resonance develops if the gap is a conventional $s^{++}$
structure.

The peak in the dynamical spin susceptibility has been observed
below $T_c$   at the antiferromagnetic wave vector {\bf Q} in
several FeSCs~ \cite{lumsden,dai,hinkov} and has been regarded by
many as the strong, yet indirect experimental evidence in favor of $s^{\pm}$
gap. [Other indirect evidences for $s^{\pm}$ SC are
the observation of a  magnetic field dependence of the
quadripartite interference peaks in STM \cite{hanaguri}
 and the very presence of the co-existence phase between antiferromagnetism and SC~\cite{fernandes}].
   These experimental findings are  particularly important because
direct phase sensitive measurements of the gap in FeSCs are lacking.

Another interpretation of the magnetic peak was put
forward by the supporters of $s^{++}$ pairing~\cite{onari_res}.  They
argued that, even in an $s^{++}$ superconductor, there is a
redistribution of a magnetic spectral weight below $T_c$ due to the
opening of a spin gap. This by itself leads
to the development of a peak in the differential dynamical spin
susceptibility (the one below $T_c$ minus the one at $T_c$).
This peak is not a resonance and occurs above $2\Delta$, contrary to the resonance which is a true bound
state and as such must be located below $2\Delta$. In principle,
this difference  should be sufficient to determine which scenario
is consistent with the data, particularly given that the gaps have
been measured by ARPES.  The difficulty, however, is that FeSCs
have several gaps of different magnitudes, and the  measured position of the
peak in the magnetic response is below $2 \Delta_{max}$ but above
$2\Delta_{min}$ and hence can  be interpreted both ways.

In this paper, we analyze in detail other properties of the spin
resonance peak in FeSCs which could help establish more accurately
whether or not the development of the peak in the spin response below
$T_c$ is the evidence for $s^{\pm}$ gap.  For this, we explore in
detail similarities and
differences  between the resonance in FeSCs and the resonance in d-wave cuprate superconductors.

The spin resonance in the cuprates has been studied in great
detail over the last 15 years. The FS in the cuprates is large and
both ${\bf k}_{F}$ and ${\bf k}_{F} + {\bf Q}$ are on the same FS
sheet (these special points are called hot spots). The velocities
at the two hot spots separated by ${\bf Q}$ are neither parallel
not antiparallel, such that the original FS and the shadow one,
with ${\bf k}$ shifted by ${\bf Q}$, simply cross at a hot spot.
In this situation,  the sign change of the gap between the two hot
spots separated by ${\bf Q}$ implies that the imaginary part of
the bare particle-hole susceptibility Im$\chi_0({\bf Q}, \Omega)$
jumps at $2 \Delta$ from its zero value below $2\Delta$ to a
finite value immediately above $2\Delta$. By Kramers-Kronig
relation, the real part of the particle-hole susceptibility
Re$\chi_0({\bf Q}, \Omega)$ then diverges logarithmically at
$2\Delta$. Because  Re$\chi_0({\bf Q}, 0)$  changes only little
between the normal and the SC state and Re$\chi_0({\bf Q},
\Omega)$ diverges upon approaching $2\Delta$ from below, the full
susceptibility  $\chi({\bf Q}, \Omega) \propto (1 - U \chi_0({\bf
Q}, \Omega))^{-1}$ develops a $\delta-$functional resonance at
some $\Omega_{res} ({\bf Q}) < 2\Delta$, where Im$\chi_0({\bf
Q},\Omega) =0$.  Upon momentum deviations from ${\bf Q}$,
 $\chi_0 ({\bf q} + {\bf Q}, \Omega)$ decreases chiefly due to the momentum
variation of the d-wave gap.  The resonance frequency approaches
zero at ${\bf q} + {\bf Q}$ equal to the distance between the two opposite points
on the FS where the $d-$wave gap vanishes. [At even larger ${\bf q}$ the new, second resonance develops and
the resonance frequency bounces back~\cite{second_res}.]

In FeSCs, the physics behind the resonance is similar to the one in the cuprates in the sense
that the sign change of the gap between hole and electron FSs
again gives rise to the divergence of the particle-hole
susceptibility at $2\Delta$. But there are also two important
differences. First, an $s-$wave gap does not induce a
downward dispersion of the resonance. Second, the resonance
in FeSCs is more likely to become incommensurate either upon doping or even
 at zero doping.  This happens because the
geometry of electron and hole FSs in FeSCs is such that there
exist special momenta ${\bf q} \neq 0$, at which a hole FS
shifted by ${\bf q} + {\bf Q}$ just touches an electron FS. At
these special ${\bf q}$, the real part of the particle-hole
susceptibility diverges as $(2\Delta-\Omega)^{-1/4}$, {\it i.e.} stronger than logarithmically.
These ${\bf q}$ appear at a finite doping if all FSs are approximated by
circles and  exist already at zero doping if we treat electron FSs
as elliptical. Because the stronger is the divergence, the farther
down is the resonance from $2\Delta$, the position of the lowest $\Omega_{res}$
shifts, upon increasing doping or ellipticity, from a
commensurate ${\bf Q}$ to an incommensurate ${\bf q} + {\bf Q}$, with a non-zero ${\bf q}$.
We show that the intensity of the resonance does not follow this shift instantly, but eventually
the intensity also becomes the strongest at an incommensurate ${\bf q}+ {\bf Q}$. A transformation
from a commensurate to an incommensurate spin resonance upon
doping has been found in the recent neutron study of the evolution of the resonance in Ba$_{1-x}$K$_x$Fe$_2$As$_2$ with doping~\cite{osborn}.  An
incommensurate spin resonance has been observed in Ref. \cite{argyriou} in a
superconducting FeTe$_{0.6}$Se$_{0.4}$.~\cite{comm}

We also discuss in detail the width of the resonance peak in
FeSCs. Experimentally, the peak is quite broad already at low $T$,
which was argued~\cite{onari_res} to be inconsistent with a true
resonance. We show that there are two effects, specific to FeSCs,
which  lead to additional broadening of the peak. First is the symmetry-imposed
 $\cos 2 \theta$  angle variation of the gap along electron FSs, the second is  pair-breaking effect on
$s^{\pm}$ superconductivity from non-magnetic impurities which
scatter between hole and electron FSs.

We  show that the impurity scattering
broadens the resonance peak, but its  position, i.e.
$\Omega_{res}$, remains essentially unaffected. Given that T$_c$ decreases
by impurity scattering, the ratio $\Omega_{res}/T_{c}$ increases
with increasing impurity concentration which correlates with
the increase of doping. This trend is in agreement with recent INS
experiments~\cite{osborn} which show that $\Omega_{res}/T_{c}$ moves up with
increasing doping.

The two key conclusions of our study is that (i) the doping
dependence and the shape of the spin resonance in FeSCs is fully
consistent with the $s^{\pm}$ gap structure, and (ii) the spin
response of an $s^{\pm}$ SC can be fully understood within an
itinerant approach, without invoking any localized moments.

The paper is organized as follows. In Sec.~\ref{sec:2} we
introduce the model and discuss the computation of a spin
susceptibility in a multi-band superconductor. In Sec. \ref{sec:3}
we discuss the doping dependence of the resonance, doping-induced
incommensuration, the role of ellipticity of the electron pockets,
and the role of the angle variation of the gap.  In Sec.
\ref{sec:4} we discuss how the resonance is affected by
non-magnetic impurities, which are pair-breaking for $s^{\pm}$
gap. Results with magnetic impurities are also included.
 Our conclusions are presented in Sec.
\ref{sec:6}

For completeness, in the Appendix we discuss one particular aspect of
the interplay between magnetic responses at ${\bf Q}$  in $s^{++}$
and $s^{\pm}$ superconductors. Namely, previous calculations have
demonstrated that there is no resonance in an $s^{++}$ superconductor,
 when the original and the shadow FSs cross. We verify what happens
 in an $s^{++}$ SC if the original and the shadow FSs do
not cross, which is the case when  both hole and electron  FSs are
near-circular and the doping is finite, such that one FS is larger
than the other. In this situation, the onset of particle hole
continuum in Im$\chi_0 ({\bf Q}, \Omega)$ shifts upwards from
$2\Delta$. We find that, even in this case, there is no resonance
for $s^{++}$ gap (and there is the resonance for $s^{\pm}$ gap).
The only feature that emerges for $s^{++}$ gap is a broad maximum
in $Im \chi ({\bf Q}, \Omega)$ above $2\Delta$.

\section{The model}
\label{sec:2}

We start from the 4-band model with two circular hole pockets at
$(0,0)$  ($\alpha$-bands) and two elliptic electron pockets at
$(\pi,0)$ and $(0,\pi)$ points in the unfolded Brillouin Zone
(UBZ) ($\beta$-bands). The quadratic part of the Hamiltonian is

\begin{eqnarray}
\label{eqH} H_2 = \sum_{\mathbf{p}, \sigma, i=1,2} \left[
\varepsilon^{\alpha_i}_{\mathbf{p}} \alpha_{i\mathbf{p}
\sigma}^\dag \alpha_{i\mathbf{p} \sigma} +
\varepsilon^{\beta_{1}}_{\mathbf{p}} \beta_{1\mathbf{p}
\sigma}^\dag \beta_{1\mathbf{p} \sigma} +
\varepsilon^{\beta_{2}}_{\mathbf{p}} \beta_{2 \mathbf{p}
\sigma}^\dag \beta_{2\mathbf{p} \sigma} \right].
\end{eqnarray}

\begin{figure}[!t]
\includegraphics[width=1.0\linewidth]{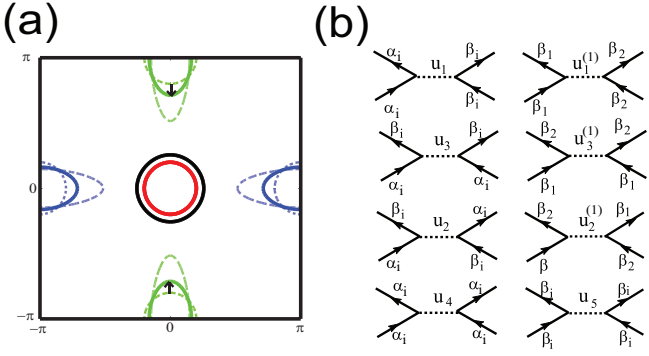}
\caption{color online) (a) Calculated Fermi surfaces for the four
band model. Solid and dashed lines for the electron FSs are for
$\epsilon=0.5$ and $\epsilon=0.8$, respectively. The arrow shows
the deviation from perfect nesting due to ellipticity; (b)
Diagrammatic representations of the interactions in the four band
model.} \label{fig1}
\end{figure}

To facilitate numerical calculations, we consider lattice dispersions for all four bands, although the resonance
essentially comes only from fermionic states near the FSs. We set
$\varepsilon^{\alpha_i}_{\mathbf{p}} = t_\alpha\left( \cos p_x
+\cos p_y \right) -\mu_i$ and $\varepsilon^{\beta_1}_{\mathbf{p}}=
\epsilon_0 + t_\beta\left( \left[1+\epsilon
\right]\cos(p_x+\pi)+\left[ 1-\epsilon \right]\cos(p_y)\right)
-\mu_1$,
 $\varepsilon^{\beta_2}_{\mathbf{p}}= \epsilon_0 +
t_\beta\left( \left[1-\epsilon \right]\cos(p_x)+\left[ 1+\epsilon
\right]\cos(p_y+\pi) \right)-\mu_1$. The momenta are measured in units of inverse Fe-Fe spacing
 $a_x = a_y =a$.
 The parameter $\epsilon$
accounts for the ellipticity of the electron pockets. To make
qualitative comparisons to experiments, we set Fermi
velocities and Fermi surfaces to be equal to those in  Refs.
\cite{LDA,Ding} For this we take $t_\alpha=0.85 eV$,
$t_\beta=-0.68 eV$, $\mu_1=1.54 eV$, $\mu_2=1.64 eV$, and
$\epsilon_0 = 0.31 eV$. For  $\epsilon=0.8$ the Fermi velocities
are $0.5eV a$ for the two degenerate $\alpha$-bands, and $v_x=0.27 eV
a$ and $v_y=0.49 eV a$ along  $x$- and $y$-directions for the
$\beta_1$-band (and vice versa for the $\beta_2$ band). The  Fermi
surfaces are shown in Fig. \ref{fig1}(a).

The interacting part of the Hamiltonian contains four-fermion
interactions with small momentum transfer and momentum transfers
$(\pi,0)$, $(0,\pi)$, and $(\pi,\pi)$.  They include $\alpha-\beta$ interactions with
momentum transfer (0, $\pi$) and ($\pi$, 0) as well as  $\beta$-$\beta$ and $\alpha$-$\alpha$ interactions with momentum transfers $(0,0)$ and $(\pi,\pi)$,
respectively, and are given by
\begin{widetext}
\begin{eqnarray}
&&H_{int}=   u_1 \sum {\alpha}^{\dagger}_{i{\bf p}_3 \sigma}
{\beta}^{\dagger}_{j{\bf p}_4 \sigma'}  {\beta}_{j{\bf p}_2
\sigma'} {\alpha}_{i{\bf p}_1 \sigma}  +  u_2 \sum
{\beta_j}^{\dagger}_{{\bf p}_3 \sigma} {\alpha}^{\dagger}_{i{\bf
p}_4 \sigma'}  {\beta}_{j{\bf p}_2 \sigma'} {\alpha}_{i{\bf p}_1
\sigma}  \nonumber \\
&& + \frac{u_3}{2} \sum \left[{\beta}^{\dagger}_{j{\bf p}_3 \sigma}
{\beta}^{\dagger}_{j{\bf p}_4 \sigma'} {\alpha}_{i{\bf p}_2
\sigma'} {\alpha}_{i{\bf p}_1 \sigma} + h.c \right] +
\frac{u_5}{2}~ \sum \left[{\alpha}^{\dagger}_{i{\bf p}_3 \sigma}
{\alpha}^{\dagger}_{i{\bf p}_4 \sigma'} {\alpha}_{i{\bf p}_2
\sigma'} {\alpha}_{i{\bf p}_1 \sigma} + {\beta}^{\dagger}_{j{\bf
p}_3 \sigma} {\beta}^{\dagger}_{j{\bf p}_4 \sigma'}
{\beta}_{j{\bf p}_2 \sigma'} {\beta}_{j{\bf p}_1 \sigma} \right] \nonumber \\
&& +  u^{(1)}_{1} \sum {\beta}^{\dagger}_{1{\bf p}_3 \sigma}
 {\beta}^{\dagger}_{2{\bf p}_4 \sigma'}  {\beta}_{2{\bf p}_2 \sigma'} {\beta}_{1{\bf p}_1 \sigma}  + u^{(1)}_{2} \sum {\beta}^{\dagger}_{2{\bf p}_3 \sigma}
 {\beta}^{\dagger}_{1{\bf p}_4 \sigma'}  {\beta}_{2{\bf p}_2 \sigma'} {\beta}_{1{\bf p}_1 \sigma}  \nonumber\\
 && + \frac{u^{(1)}_{3}}{2}  \sum \left[ {\beta}^{\dagger}_{2{\bf p}_3 \sigma}
 {\beta}^{\dagger}_{2{\bf p}_4 \sigma'}  {\beta}_{1{\bf p}_2 \sigma'} {\beta}_{1{\bf p}_1 \sigma}  + h.c. \right]   + u^{(2)}_{1} \sum {\alpha}^{\dagger}_{1{\bf p}_3 \sigma}
 {\alpha}^{\dagger}_{2{\bf p}_4 \sigma'}  {\alpha}_{2{\bf p}_2 \sigma'} {\alpha}_{1{\bf p}_1 \sigma} \nonumber\\
&&   + u^{(2)}_{2} \sum {\alpha}^{\dagger}_{2{\bf p}_3 \sigma}
 {\alpha}^{\dagger}_{1{\bf p}_4 \sigma'}  {\alpha}_{2{\bf p}_2 \sigma'} {\alpha}_{1{\bf p}_1 \sigma} + \frac{u^{(2)}_{3}}{2} \sum \left[ {\alpha}^{\dagger}_{2{\bf p}_3 \sigma}
 {\alpha}^{\dagger}_{2{\bf p}_4 \sigma'}  {\alpha}_{1{\bf p}_2 \sigma'} {\alpha}_{1{\bf p}_1 \sigma}  + h.c. \right] \quad.
 \label{eq:2}
\end{eqnarray}
\end{widetext}
The  vertices are shown
diagrammatically in Fig. \ref{fig1}(b).
For simplicity, we approximate all interactions as angle-independent, i.e., neglect the angle dependence introduced by dressing the
interactions by coherence factors associated with the hybridization of Fe $d-$orbitals. These coherent factors do play a role for the structure of the SC gap~\cite{saurabh}, but do not substantially modify the spin resonance~\cite{maier}.

In the magnetically-disordered state transverse and longitudinal components of the spin susceptibility are undistinguishable, and we focus below on the transverse part. In the matrix RPA approximation, which we adopt, the transverse components of the full spin susceptibility
$\chi^{i,j}$, where $i$ and $j$ are band indices, are related to transverse components of the bare
 susceptibility $\chi^{i,j}_{0}$  as
 \begin{equation}
\chi^{i,j}=\chi^{i,j}_{0}+\chi^{i,j^{\prime}}_{0}
 u_{i^\prime,j^\prime}  \chi^{i^{\prime},j} \quad.
\label{n_1}
\end{equation}
 The summation over repeated band indices is assumed and $u^{i^\prime j^\prime}$ are matrix elements of the interactions shown in Fig.\ref{fig1}(b) (contributions from all $u^k$ are included in (\ref{n_1})). The solution of Eq.(\ref{n_1}) in matrix form is straightforward:
$\hat\chi=\hat\chi_{0}(1-{\hat u} {\hat\chi}_{0})^{-1}$.

The components of the bare spin susceptibility $\hat\chi_{0} =
\chi^{i,j}_0 ({\bf q},{\rm i} \Omega_m)$ are given by usual
combinations of normal and anomalous Green's functions
\begin{eqnarray}
\chi^{ij}({\bf p},{\rm i} \Omega_m)&=&- \frac{T}{2N} \sum_{{\bf
k}, \omega_n} {\rm Tr}
\left[ G^i({\bf k + p}, {\rm i} \omega_n + {\rm i} \Omega_m) G^j({\bf k} , {\rm i} \omega_n) \right. \nonumber \\
&+& \left. F^i({\bf k + p}, {\rm i} \omega_n + {\rm i} \Omega_m)
F^j({\bf k} , {\rm i} \omega_n)\right] \label{eq:bare_chi}
\end{eqnarray}
where $G^i({\bf
k},i\omega_n)=-\frac{i\omega_n+\varepsilon^{i}_{\bf
k}}{\omega^2_n+(\varepsilon^{i}_{\bf k})^2+(\Delta_{\bf k
}^{i})^{2}}$ and $F^i({\bf k},i\omega_n) = \frac{\Delta_{\bf k
}^{i}}{\Omega^2_n+(\varepsilon^{i}_{\bf k})^2+(\Delta_{\bf k
}^{i})^{2}}$.

The main contribution to the full spin susceptibility with momenta
near ${\bf Q}$  comes from susceptibilities and interactions which
involve hole and electron states (particle-hole bubbles made of
one electron and one hole propagator, and $u_1$ and $u_3$
interaction terms in Eq. (\ref{eq:2}), see Ref. \cite{chubukov}).
For completeness, in numerical calculations we will keep all terms
in the matrix equation for the full susceptibility. The results
for the full $\sum_{ij} \chi^{ij}({\bf p},{\rm i} \Omega_m)$
obtained this way do not differ much from those obtained by
keeping only electron-hole bubbles and $u_1$ and $u_3$ interaction
terms.

We do not present explicitly interactions  leading to
superconductivity and not discuss the solution of the pairing
problem. This has been done in numerous other works on this
subject~\cite{saurabh}. We take the results of these studies as
input and set the gap to be of $s^{\pm}$ type, with different sign
between hole and electron pockets. More specifically, we set the
gap to be a constant along the hole FSs, $\Delta^{\alpha_{1,2}} =
\Delta_h$, and set the gap along the two electron FSs to be
 $\Delta^{\beta_{1,2}} = \Delta_e (1 \pm r
\cos 2 \phi)$, where $\phi$ is the angle counted from $k_x$
direction in the UBZ.  The gap $\Delta^{\beta_{1,2}}$ has no nodes
if $r<1$ and has "accidental" nodes when $r>1$ at non-symmetry
selected directions $\cos 2\phi =1/r$. The numerical results
presented below are obtained for  $u_{1} = u_3 \approx 0.25$eV (
these numbers guarantee that system remains paramagnetic in the
normal state), and $u_5=u_2=0.5u_1$, $u^{(j)}_{i}=0.1u_i$. For
simplicity we set $\Delta_h = -\Delta_e = \Delta$ although
$\Delta_h$ and $- \Delta^{\beta_{1,2}} (\phi = \pi/4)$ do not have
to be equal. For the gap we used $\Delta=0.02t_\alpha$.  For
better convergence of numerical series we added a small damping
$\Gamma=1$meV to fermionic dispersion.

\section{The results}
\label{sec:3}

We present results systematically in three installations. We first consider the
doping evolution of the resonance for a simple plus-minus gap and circular FSs, then we include
 the ellipticity of electron pockets,  and finally we also  include  the angular dependence of the SC gap.
 For consistency with the experiments, we show all results in the folded BZ, when the commensurate resonance is at ${\bf Q} = (\pi,\pi)$.

\subsection{Doping dependence of the resonance for plus-minus gap and circular FSs}

This case is captured by setting $\epsilon=0$ and $r=0$ in $\Delta^{\beta_{1,2}}$.  The doping dependence is parameterized  by
the change in the chemical potential $\delta \mu$.

\begin{figure}[!t]
\includegraphics[width=1.0\linewidth]{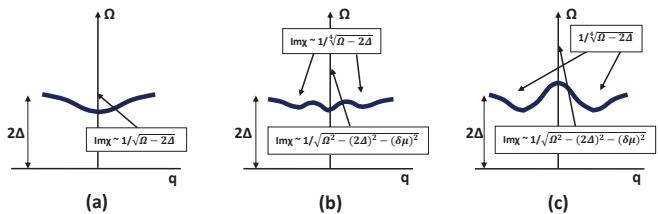}
\caption{(color online) Schematic form of the momentum dispersion
of the spin resonance frequency $\Omega_{res}$ (solid line)
  at various doping and at zero ellipticity.  Red dashed region is
 of the particle-hole continuum.  The formulas show how the bare spin susceptibility diverges at the bottom of the
 continuum. Panel a -- zero doping, panels (b),(c) -- a finite doping (larger for c).}
\label{fig2schematic}
\end{figure}

Much in this case can be understood analytically.  At zero doping,
$\chi_0 ({\bf Q}, \Omega)$ has a strong, square-root singularity
at $\Omega = 2\Delta$. $Im \chi_0 ({\bf Q}, \Omega)$ diverges as
$1/\sqrt{\Omega -2\Delta}$ at $\Omega > 2\Delta$ and $Re \chi_0
({\bf Q}, \Omega)$ diverges as $1/\sqrt{\Omega -2\Delta}$ at
$\Omega < 2\Delta$). For incommensurate momenta, $Im \chi_0 ({\bf
q} + {\bf Q}, \Omega)$ undergoes a finite jump at $\Omega =
2\Delta$ and $Re \chi_0 ({\bf q} + {\bf Q}, 2\Delta)$ diverges
logarithmically, as long as $q < 2k_F$. In this situation, the
lowest $\Omega_{res}$ and the largest intensity are at the
commensurate momentum ${\bf Q}$. We illustrate this in Fig.
\ref{fig2schematic}(a).
At a finite doping, $\chi_0 ({\bf Q}, \Omega)$ still have a
square-root singularity, but now the lower boundary of the
particle-hole continuum shifts up and the singularity is located
at a larger $\Omega_{\bf Q} = \sqrt{(2\Delta)^2+\delta\mu^2}$. The
resonance in the full $\chi ({\bf Q}, \Omega)$ is located below
$\Omega_{\bf Q}$, but because $\Omega_{\bf Q}$ increases with
$\delta \mu$, the position of the resonance at the commensurate
${\bf Q}$ also increases.  Meanwhile, there exists the range of
incommensurate momenta ${\bf q} + {\bf Q}$ for which the bottom of
the particle-hole continuum is still located at $2\Delta_0$. These
are $q_{min} < q < q_{max}$, where $q_{min} = \sqrt{k^2_F +
\delta\mu^2} - k_F$ and $q_{max}=\sqrt{k^2_F + \delta\mu^2} +
k_F$.  For $q$ inside this range, $Im \chi_0 ({\bf q} + {\bf Q},
\Omega)$ undergoes a finite jump at $\Omega = 2\Delta$ and $Re
\chi_0 ({\bf q} + {\bf Q}, 2\Delta)$ diverges logarithmically, as
before, but at the end points, i.e., at $q = q_{min}$ and $q =
q_{max}$, hole and electron FSs touch each other after a shift by
${\bf Q}$, and $\chi_0 ({\bf q} + {\bf Q}, \Omega)$ diverges by a
power-law, this time as $1/(2\Delta -\Omega)^{1/4}$. Because of
stronger divergence, the resonance in the full $\chi ({\bf q} +
{\bf Q}, \Omega)$ at these $q$ shifts down from $2\Delta$ more
than for other $q$, hence the dispersion of the resonance develops
minima at $q = q_{min}$ and $q= q_{max}$.

At small doping, these two minima at incommensurate $q$ are local
minima because $\Omega_{res} ({\bf q} + {\bf Q})$  had an upward
dispersion around ${\bf Q}$ at zero doping, one needs some finite
doping to change the sign of the slope at $q=0$.  As a result, at
small but finite doping, the dispersion of the resonance has the
global minimum at $q=0$ and two roton-like minima at $q_{min}$ and
$q_{max}$ (see Fig.\ref{fig2schematic}(b)).  Once the doping gets larger, the
resonance energy at $q=0$ keeps going up together with the bottom
of the particle-hole continuum, and eventually the global minimum
of the resonance dispersion discontinuously shifts to a finite
$q$, i.e., to an incommensurate momentum  (see Fig.\ref{fig2schematic}(c)).

\begin{figure}[!t]
\includegraphics[width=1.0\linewidth]{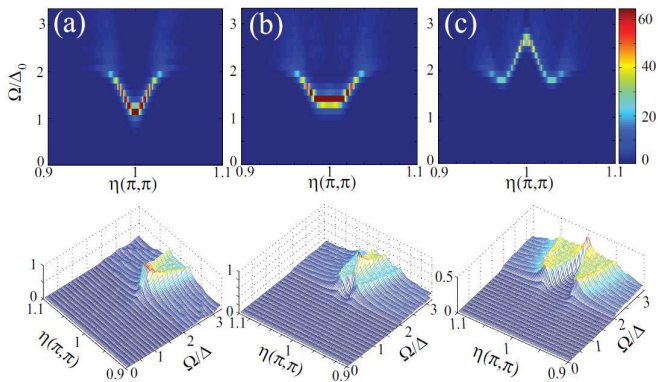}
\caption{(color online) The imaginary part of the total physical
susceptibility Im$\chi({\bf q+Q},\Omega)$ for various doping
levels at zero ellipticity: (a) $x=0$,
(b)$x=0.01$($\delta\mu=-0.02$), and
(c)$x=0.03$($\delta\mu=-0.051$). The lower panel shows the
corresponding behavior of the imaginary part of the bare spin
susceptibility.} \label{fig2}
\end{figure}

We reproduced this behavior in the numerical analysis. We display
the numerical results in Fig. \ref{fig2}. The upper panels show
the dispersion of the resonance energy, the lower panels show the
imaginary part of $\chi_0 ({\bf q} + {\bf Q}, \Omega)$.
Fig.\ref{fig2}(a) shows the behavior at zero doping. The
`square-root' divergence of $Im  \chi_0 ({\bf Q}, \Omega)$ at
$\Omega=2\Delta$ is clearly visible, while for nonzero
$q<q_{max}=2k_F$ we observe a `jump' in  $Im  \chi_0 ({\bf q} +
{\bf Q}, \Omega)$ at $\Omega=2\Delta$, in agreement with
analytical consideration. We also see, at frequencies above
$2\Delta$, the "upward" dispersion of the maximum in $Im  \chi_0
({\bf q} + {\bf Q}, \Omega)$.  This dispersion originates from the
momentum dependence of the static $\chi_0 ({\bf q} + {\bf Q})$ and
gives rise to the upward dispersion of $\Omega_{res} ({\bf q} +
{\bf Q})$ in the upper panel of Fig. \ref{fig2}(a).

In Figs. \ref{fig2}(b) we show the behavior of $Im \chi_0$ and the
dispersion of the resonance at small but finite dopings $x=0.01$
($\delta \mu = - 0.02$). We again see the square root singularity
in  $Im  \chi_0 ({\bf q} + {\bf Q}, \Omega)$  at the commensurate
momentum $q=0$, but now it shifts to a frequency above $2\Delta$,
consistent with analytical
$\Omega=\sqrt{(2\Delta)^2+\delta\mu^2}$. We also clearly see
additional singularity in $Im  \chi_0 ({\bf q} + {\bf Q}, \Omega)$
at an incommensurate momentum, which we associate with $q_{min}$
(other momentum, $g_{max}$ is outside the momentum range in
\ref{fig2} and $Im  \chi_0 ({\bf q} + {\bf Q}, \Omega)$ at such
${\bf q}$ already quite small).  The resonance frequency
$\Omega_{res} ({\bf q} + {\bf Q})$ still has a minimum at ${\bf q}
=0$, but because of an additional singularity in $Im  \chi_0 ({\bf
q} + {\bf Q}, \Omega)$ at ${\bf q}={\bf q}_{min}$, the dispersion
of $\Omega_{res}$ around the minimum becomes rather flat. As
doping is further increased to $x=0.03$ ($\delta\mu=-0.051$)
(Fig.\ref{fig2}(c)), the square root singularity at ${\bf q} =0$
is pushed further outward and $\Omega_{res} ({\bf Q})$ eventually
moves to above $2\Delta$, while the one at ${\bf q} = {\bf
q}_{min}$ stays below $2\Delta$. As a result, the minimum of
$\Omega_{res}$ shifts from ${\bf q}=0$ to ${\bf q}={\bf q}_{min}$,
as is clearly seen in the upper panel of Fig.\ref{fig2}(c).

We found that the maximal intensity of the resonance does not
immediately follow the position of the minimum of $\Omega_{res}$
and over some range of dopings is still the largest at the
commensurate momentum ${\bf Q}$ even when the minimum of
$\Omega_{res}$ already shifts to an incommensurate ${\bf q} + {\bf
Q}$. But as doping increases even further, the maximum of the
intensity eventually shifts to an incommensurate momentum.

\subsection{Role of ellipticity of the electron pockets.}

\begin{figure}[!t]
\includegraphics[width=1.0\linewidth]{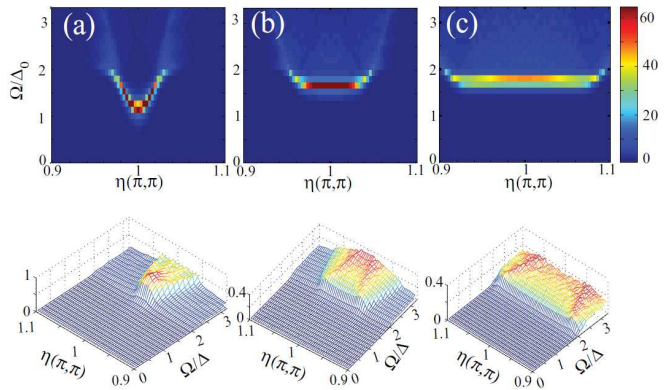}
\caption{(color online) The imaginary part of the total physical
susceptibility Im$\chi({\bf q+Q},\Omega)$ for various degrees of
ellipticity at zero doping: (a) $\epsilon=0.1$, (b)$\epsilon=0.3$,
and (c)$\epsilon=0.5$.  The lower panel shows the behavior of the
imaginary part of the bare spin susceptibility.} \label{fig2n}
\end{figure}

We next consider the case of elliptical electron pockets, when $\epsilon$ is nonzero.
In this case, resonance may become incommensurate already at zero doping. Indeed, at
$x=0$ and non-zero ellipticity,  Im $\chi_0 ({\bf Q},\Omega)$ has a finite jump at $2\Delta$
because hole and electron velocities at each of the  four hot spots (points where a hole pocket shifted by
${\bf Q}$ crosses an electron pocket) are neither parallel nor antiparallel. In other words,
a square-root singularity at ${\bf Q}$ is cut by $\epsilon$ and is replaced by a jump.
However, the power-law, $1/(2\Delta -\Omega)^{1/4}$, singularities at incommensurate $q_{min}$ and $q_{max}$
 survive. Like before, ${\bf q}_{min}$ and ${\bf q}_{max}$ are the points where the hole FS, shifted by {\bf Q}, touches an elliptical
electron FS. At the smaller ${\bf q} = {\bf q}_{min}$  four hot
spots transform into two,  at a larger ${\bf q} = {\bf q}_{max}$,
the remaining two hot spots disappear.  Because  the shift of
$\Omega_{res}$ from $2\Delta$  is larger, the stronger is the
divergence of   Im $\chi_0 ({\bf Q},\Omega)$ at $\Omega =
2\Delta$, the minimum of  the dispersion of $\Omega_{res}$ should
eventually shift to an incommensurate momentum once ellipticity
becomes large enough to overcome a positive curvature of the
dispersion near ${\bf Q}$ which persists up to some finite
$\epsilon$. We found the same trends in the  numerical
calculations. In particular, in Fig.\ref{fig2n} we show  the
evolution of the dispersion of the resonance and the form of $Im
\chi_0 ({\bf q} + {\bf Q}, \Omega)$ as a function of ellipticity
at zero doping. We clearly see that once $\epsilon$ increases, the
dispersion becomes more flat near the minimum at ${\bf Q}$, and
eventually the minimum of the dispersion shifts to a non-zero
${\bf q}$, i.e., to an incommensurate momentum.

\begin{figure}[!t]
\includegraphics[width=1.0\linewidth]{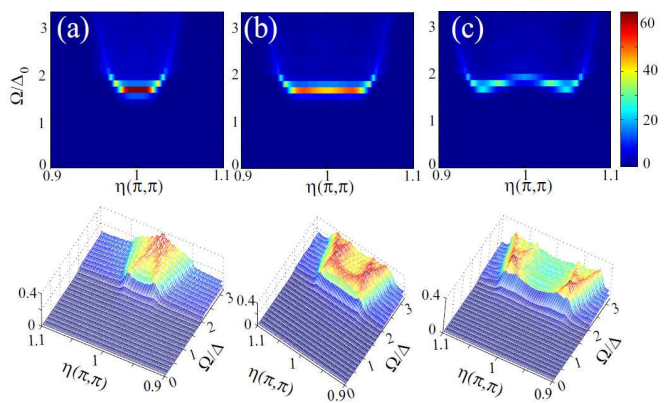}
\caption{(color online) The imaginary part of the total physical
susceptibility Im$\chi({\bf q+Q},\Omega)$ for various doping
levels [(a) $x=0$($\delta\mu=0$), (b)$x=0.02$($\delta\mu=-0.035$),
and (c)$x=0.05$($\delta\mu=-0.076$) and $\epsilon = 0.3$. The
lower panel shows the corresponding behavior of the imaginary part
of the bare spin susceptibility.} \label{fig3n}
\end{figure}

In Fig. \ref{fig3n} we combine finite doping and finite
ellipticity.  We see the same tends as before, namely at a given
non-zero $\epsilon$ (we set $\epsilon =0.3$ for definiteness), the
resonance is  quite broad already at $x=0$ (Fig. \ref{fig3n}(a)),
it gets even more broad as $x$ increases (Fig. \ref{fig3n}(b)),
and eventually incommensurate minima appear at large enough $x$
(Fig. \ref{fig3n}(c)).  Note that at non-zero ellipticity the
development of incommensurate minima in $\Omega_{res}$ at is less
pronounced effect than at $\epsilon =0$ and the key effect at
small doping is the broadening of the resonance peak. This is
because the bottom of the particle-hole continuum at ${\bf Q}$
does not move up with doping as long as an electron ellipses and a
hole circle shifted  by ${\bf Q}$  cross, and $\Omega_{res}$ at
${\bf Q}$ and at ${\bf q} + {\bf Q}$ with $q = q_{min}$ remain
near-equal. Still, at larger dopings the resonance definitely
becomes incommensurate, i.e., the minimum of $\Omega_{res}$ and
the largest intensity shift to a incommensurate position ${\bf q}
= {\bf q}_{min}$.  There is another local minimum at ${\bf q} =
{\bf q}_{max}$, like at $\epsilon =0$, but that one corresponds to
large enough ${\bf q}$ where $Im \chi_0$ is already small.

\subsection{Role of the gap anisotropy}

\begin{figure}[!t]
\includegraphics[width=1.0\linewidth]{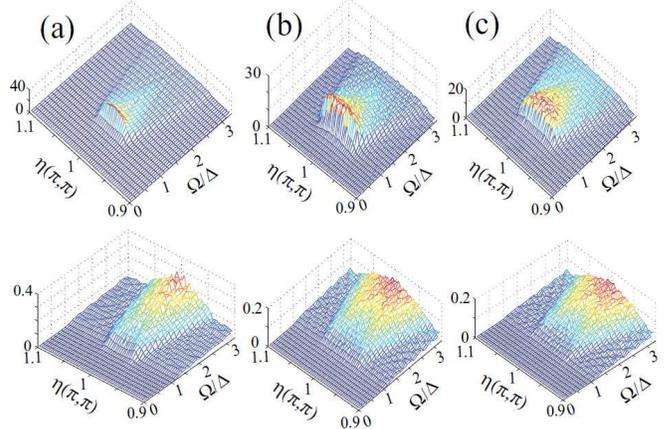}
\caption{(color online) The imaginary part of the bare interband
susceptibility Im$\chi^{\alpha_1 \beta_1}({\bf Q},\omega)$(lower
panel) and the total physical susceptibility (upper panel) for
(a)$r=0.5$, (b)$r=1$, and (c)$r=2$. We set $x=0.03$  and
$\epsilon=0.5$.} \label{fig5new}
\end{figure}

Another interesting aspect of the physics of the spin resonance in
FeSCs is the role played by the angular dependence of the
s$^{+-}$-wave gap. We remind that in FeSCs the gaps on the two
electron FSs behave as $\Delta^{\beta_{1,2}} = -\Delta (1 \pm r
\cos 2 \phi)$, and $r$ is generally a finite number. On general
grounds we expect that the resonance gets broader simply because
$2\Delta$ becomes ``soft" variable once the  magnitude of
$2\Delta$ varies along the electron FS. Only at $T=0$ and in the
idealized case of no impurity scattering this ``softness" does not
matter because only fermions in the immediate vicinity of hot
spots contribute to singularity in $Im \chi_0 ({\bf q} + {\bf Q},
\Omega)$, and these fermions have some fixed $\Delta_k$ and
$\Delta_{k+ {\bf q} +{\bf Q}}$.  At a finite $T$ and/or in the
presence of some residual scattering, the momentum range for
fermions contributing to the resonance increases and the angular
dependence of the electron gap becomes relevant. Numerical
calculations are in line with this reasoning. In Fig.\ref{fig5new}
(lower panel) we show the evolution with $r$ of the bare
susceptibility  at ${\bf Q}$ taken between $\alpha_1$ and
$\beta_1$ bands.  Other interband susceptibilities show
qualitatively similar behavior. We set ellipticity and doping to
be non-zero ($\epsilon = 0.5$, $x =0.03$), so there is no
power-law singularity at ${\bf Q}$. For $r=0$, Im
$\chi^{\alpha_1,\beta_1}_0 ({\bf Q}, \Omega)$ then has a finite
jump at $2\Delta$. We see that the jump persists also at a finite
$r$, but its magnitude is reduced and the jump essentially
disappears at $r > 1$.  The reduction of the jump affects the
behavior of the full spin susceptibility, which we show in Fig.
\ref{fig5new} (upper panel). We see that the resonance peak, which
is already quite broad at $r=0$, become even more broad at  a
finite $r$, and at large enough $r$ the width of the resonance
peak is essentially controlled by $r$.

The broadening of the peak at ${\bf Q}$ is wider than in the
cuprates because there the FS is large and the momenta which
contribute to the resonance are confined to near vicinity of hot
spots, where $\Delta_k$ and $\Delta_{k+ {\bf Q}}$ are rather flat
and can be treated as constants~\cite{second_res}. In
FeSCs, the FSs are much smaller and, accordingly, the effect of
gap variation on the width of the resonance is stronger.

Strong enhancement of the width of the resonance peak due to angle
variation of the SC gap is the possible explanation why the
observed peak at ${\bf Q}$ is wider than one should expect from
the fully gapped $s^{\pm}$-state~\cite{onari}. Another possible
explanation is the effect of impurities, which we consider in the
next section.

The effect of the gap variation on the dispersion of the resonance
is strongest at ${\bf Q}$ because for incommensurate momenta and,
particularly, for $q = q_{min}$, the gaps on hole and electron FSs
differ quite substantially by magnitude, and, when $r
>1$, have the same sign.  A simple reasoning shows that, in this
situation, two effects happen: the bottom of the particle-hole
continuum goes down near ${\bf q}_{min}$, and, at the same time, the
intensity of the resonance peak at ${\bf q}_{min}$ drops, and for $r>1$
the resonance at ${\bf q} \sim q_{min}$  simply does not exist. The
outcome is that, as $r$ increases, the resonance dispersion
flattens up near ${\bf Q}$ and the intensity drops away from ${\bf
Q}$. As shown in Fig.\ref{fig5new} the numerical analysis is again in line with this reasoning.

\section{Effect of Impurities}
\label{sec:4}

In this section we study how the position and the width of spin
resonance are affected by elastic impurity scattering.  We
consider both magnetic and non-magnetic impurities. To keep the
calculations tractable we
make several simplifying assumptions: (i) we only consider
resonance at $T=0$ and at the commensurate momentum ${\bf Q}$,
(ii) we neglect ellipticity of electron pockets and the angle
variation of the SC gap, and (iii) we approximate the band
dispersion by the two-band model of one hole and one electron
pockets ($\alpha$- and $\beta$-fermions) separated by ${\bf Q} =
(\pi,\pi)$ and perfectly nested, i.e., set
$\varepsilon^{\alpha}({\bf k})  =  -\varepsilon^{\beta}({\bf
k}+{\bf Q})$ and, in the absence of impurities, $\Delta^\alpha =
-\Delta^\beta = \Delta$.   We show that impurity scattering
broadens the resonance but its energy position $\Omega_{res} ({\bf
Q})$ can remain essentially intact.  Because $T_c$ for $s^{\pm}$
pairing drops with impurity scattering, the ratio
$\Omega_{res}/T_c$ increases, in agreement with the experimental
data. Because both the broadening of the resonance and the
increase of $\Omega_{res}/T_c$ with impurity scattering are
related to $s^{\pm}$ gap symmetry, we expect that these two
results survive for more realistic 4-5 band models with elliptical
electron FSs and angle-dependent gap.

\subsection{Method}

Non-magnetic impurities  affect electrons via the impurity scattering potential in the form
\begin{widetext}
\begin{equation}\label{H_imp}
H_{imp}=\sum_{{\bf k_{1,2}}}\left[ U_0({\bf k}_1-{\bf k}_2)\left(
\alpha_{{{\bf k}_1}\sigma}^{\dagger}\delta_{\sigma
\sigma^{\prime}}\alpha_{{{\bf k}_2}\sigma^{\prime}} + \beta_{{{\bf
k}_1}\sigma}^{\dagger}\delta_{\sigma \sigma^{\prime}}\beta_{{{\bf
k}_2}\sigma}\right)\,+\, U_{\pi}({\bf k}_1-{\bf k}_2)\left(
\alpha_{{{\bf k}_1}\sigma}^{\dagger}\delta_{\sigma
\sigma^{\prime}}\beta_{{{\bf k}_2}\sigma^{\prime}} + \beta_{{{\bf
k}_1}\sigma}^{\dagger}\delta_{\sigma \sigma^{\prime}}\alpha_{{{\bf
k}_2}\sigma^{\prime}} \right) \right]
\end{equation}
\end{widetext}
where $U_0 ({\bf k_1 -k_2})$ and $U_{\pi}({\bf k_1 -k_2})$ are
intra- and inter-band scattering terms, respectively. For $U_0$,
$\bf k_1$ and $\bf k_2$ are both near $(0,0)$ or ${\bf Q}$, while
for $U_{\pi}$ $\bf k_1 \approx (0,0)$ and $\bf k_2 \approx \bf Q$
or vice versa. Quite generally one expects $U_{\pi}<U_0$ for any
impurity potential with a finite range in real space.

Paramagnetic impurities in turn affect electrons via the
spin-dependent potential
\begin{widetext}
\begin{eqnarray}\label{H_p}
H_{p\,mag} = \sum_{\bf{k_{1,2}}}\; U^p_0({\bf k_1}-{\bf
k_2})\left( \alpha_{{\bf k_1}\sigma}^{\dagger}\vec{\sigma}_{\sigma
\sigma^{\prime}}\cdot\vec{S}\alpha_{{\bf k_2}\beta} + \beta_{{\bf
k_1}\sigma}^{\dagger}\vec{\sigma}_{\sigma
\sigma^{\prime}}\cdot\vec{S}\beta_{{\bf k_2}\beta}\right)\,+ \nonumber\\
U^p_{\pi}({\bf k_1}-{\bf k_2})\left( \alpha_{{\bf
k_1}\sigma}^{\dagger}\vec{\sigma}_{\sigma
\sigma^\prime}\cdot\vec{S}\beta_{{\bf k_2}\sigma^\prime} +
\beta_{{\bf k_1}\sigma}^{\dagger}\vec{\sigma}_{\sigma
\sigma^\prime}\cdot\vec{S}\alpha_{{\bf k_2}\sigma^\prime} \right)
\end{eqnarray}
\end{widetext} where $\vec{\sigma} = (\sigma_x,\sigma_y,\sigma_z)$
and $\sigma_i$ are Pauli matrices, and $\vec{S}$ is the static
impurity spin with the property $<{
S_iS_j}>_{impurity~sites}=\frac{S(S+1)}{3} \delta_{i,j}$,
$i,j\in\{x,y,z\}$.

Fermionic propagators for an $s^{\pm}$  superconductors in the
presence of non-magnetic impurities have been considered before
(see e.g., Ref. \onlinecite{chubukov}). The extension to
include paramagnetic impurities is straightforward, and yields
\begin{eqnarray}\label{g_f_imp}
G^{\alpha,\beta} &= &- \frac{Z\,i\omega_m +
\varepsilon_{{\bf k}}^{\alpha,\beta}} {Z^2(\omega_m^2 +
|\bar{\Delta}_\omega|^2) +
(\varepsilon_{{\bf k}}^{\alpha,\beta})^2} \nonumber\\
F^{\alpha,\beta} &= & \pm\frac{Z\bar{\Delta}_\omega}{Z^2(\omega_m^2 +
|\bar{\Delta}_\omega|^2) + (\varepsilon_{{\bf k}}^{\alpha,\beta})^2}
\end{eqnarray}
with
\begin{eqnarray}\label{Z1,Delta_bar}
 Z&=&1+\frac{u_0+u_{\pi}+u^p_0+u^p_{\pi}}
{\sqrt{\omega_m^2+|\bar{\Delta}_\omega|^2}} \nonumber\\
\left( \frac{\bar{\Delta}_\omega}{\Delta_b} -1\right)^2&=& b^2\,
\frac{\bar{\Delta}^2_\omega}{\omega_m^2+|\bar{\Delta}_\omega|^2} \nonumber\\
b&=&\frac{2\left(u_{\pi}+u^p_0\right)}{\Delta_b}.
\end{eqnarray}
where we have adopted, for brevity, the notation $u_0\equiv
n_{imp}|U_0(0)|^2$ (and a similar  for $u_{\pi}, u_0^p$, and
$u_\pi^p$), with $n_{imp}$ being the impurity concentration. In
Eq. (\ref{Z1,Delta_bar}), ${\bar \Delta}_\omega$ is the actual,
frequency-dependent gap in the presence of impurity scattering,
and $\Delta_b$ is the order parameter related to ${\bar
\Delta}_\omega$ via
 \beq
 \frac{\Delta_b}{\Delta_0}  =
\frac{\int_0^{\omega_{max}}d\omega_m \frac{{\bar
\Delta}_\omega}{|{\bar\Delta}_\omega|^2 +
\omega^2_m}}{\int_0^{\omega_{max}}d\omega_m
\frac{\Delta_0}{\Delta^2_0 + \omega^2_m}}
 \label{f_1}
 \eeq
For any $b \neq 0$, a non-zero fermionic DOS $N(\omega) = (m/2\pi) Re\left[\omega/\sqrt{\omega^2 -{\bar \Delta}^2_\omega}\right]$ extends to frequencies below $2\Delta$. For  $0<b <1$, the system still preserves the gap in the fermionic DOS near zero frequency, i.e., over some range of $\omega$ near $\omega =0$, $N(\omega) =0$. For $b >1$, the system enters into a gapless regime, in which ${\bar \Delta}_\omega$ scales as $i \omega$ at the lowest frequencies and $N(\omega =0)$ is non-zero, although reduced compared to the DOS in the normal state.

Observe that the gap renormalization comes from  interband
non-magnetic impurity scattering and intraband magnetic impurity
scattering which are pair-breaking for $s^{\pm}$ superconductor
(both also reduce $T_c$). At the same time intraband non-magnetic
impurity scattering and interband magnetic impurity scattering do
not contribute to $b$ and therefore are not pair-breaking.  If
only these two scattering were present, $b$ would be zero, ${\bar
\Delta}_\omega$ would be equal to $\Delta$, and Eq. (\ref{f_1})
would then yield ${\bar \Delta}_\omega = \Delta_b = \Delta$, i.e.,
the gap would not be affected by impurities. The same analysis
extended to a finite $T$ shows that $u_0$ and $u^p_\pi$ also do
not affect $T_c$.

The next step is to use normal and anomalous Green's functions
from  (\ref{g_f_imp}) and compute spin susceptibility for a dirty
BCS superconductor.  This has to be done with care, though, as one
has to include not only self-energy corrections, incorporated in
(\ref{g_f_imp}), but also series of vertex corrections which are
of the same order as self-energy terms. We follow the treatment by
Gorkov and Rusinov \cite{bib:GR}, sum up ladder series of vertex
correction
diagrams and after straightforward algebra obtain
\begin{widetext}
\begin{equation}\label{vertex_corr_chi_mag_imp}
\chi_0({\bf Q}, \Omega) = N_0\int  \frac{d\omega}{2\pi}
F_{\Delta^2}\left[\frac{1}{1+(u_0 + u_{\pi}-\frac{1}{3}(u_0^p +
u_{\pi}^p))F_{\Delta^2}+(u_0 - u_{\pi} + \frac{1}{3}(u_0^p -
u_{\pi}^p))\frac{F^2_{\omega\Delta}}{F_{\Delta^2}}} \right].
\end{equation}
\end{widetext}
where $N_0$ is the 2D density of states, and
\begin{equation}
F_{\Delta^2} =
\frac{-\omega(\omega+\Omega)+\bar{\Delta}_{\omega}\bar{\Delta}_{\omega+\Omega}+f_{\omega}f_{\omega+\Omega}}{f_{\omega}f_{\omega+\Omega}\left(
f_{\omega}+f_{\omega+\Omega} + 2u_0 + 2u_{\pi}\right)}
\end{equation}
\begin{equation}
F_{\omega\Delta} =
\frac{\omega\bar{\Delta}_{\omega+\Omega}-(\omega+\Omega)\bar{\Delta}_{\omega}}{f_{\omega}f_{\omega+\Omega}\left(
f_{\omega}+f_{\omega+\Omega} + 2u_0 + 2u_{\pi}\right)}
\end{equation}
with $f_{\omega} = \sqrt{\bar{\Delta}^2_{\omega}-\omega^2}$.

As before, we assume that the full susceptibility $\chi ({\bf Q},
\Omega)$ is related to $\chi_0 ({\bf Q}, \Omega)$ as $\chi ({\bf
Q}, \Omega) = \chi_0 ({\bf Q}, \Omega)/(1 - u_{spin} \chi_0 ({\bf
Q}, \Omega))$.
Where $u_{spin}$ is the residual interaction in the spin channel
(see Sec\ref{sec:2}).

We solved for ${\bar \Delta}_\omega$, substituted the result into  (\ref{vertex_corr_chi_mag_imp}), integrated over frequency
and obtained bare $\chi_0({\bf Q}, \Omega)$ and the full Im $\chi ({\bf Q}, \Omega)$ as functions of impurity scattering.

\subsection{Results}

We first discuss the role of non-magnetic impurities $u_0$ and
$u_{\pi}$ and for this purpose set $u^p_0$ and $u^p_{\pi}$ to zero in Eq. (\ref{Z1,Delta_bar}).
 Both $u_0$ and $u_\pi$ renormalize quasiparticle $Z$, but only $u_{\pi}$
  contributes to pair-breaking parameter $b$.

\begin{figure}[t]
\includegraphics*[width=1.0\linewidth]{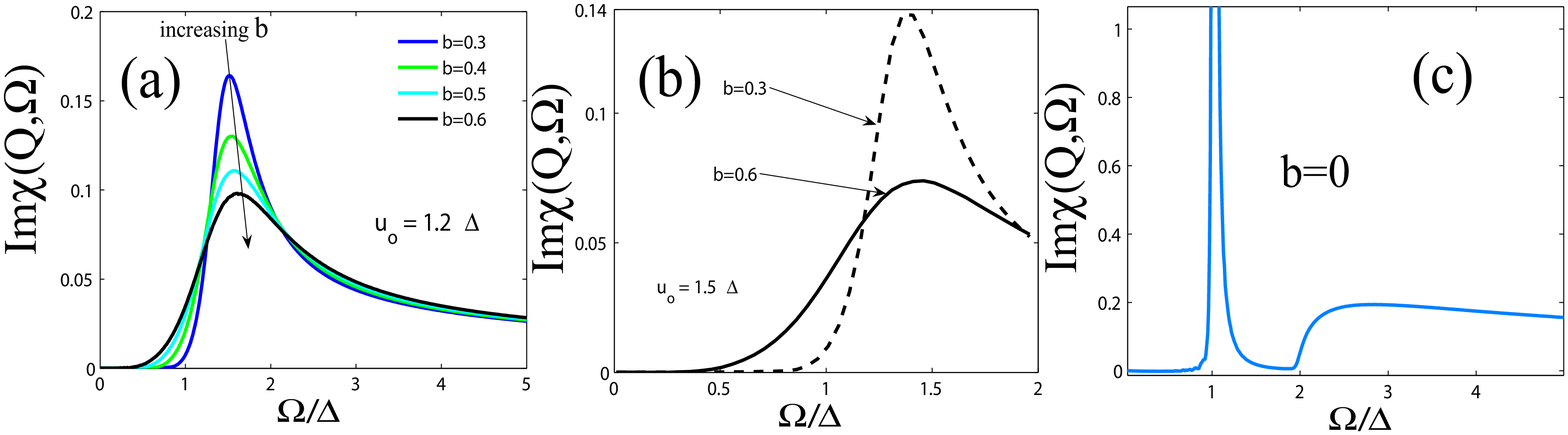}
\caption{Calculated Im$\chi({\bf Q},\Omega)$ for scattering by non-magnetic impurities,
 for different
values of $b$, which scales with the strength of the interband
impurity scattering. We used $u_0=1.2 \Delta$ (a) and
$u_0=1.5\Delta$ (b). Figure (c) shows, for comparison, the
resonance peak for $b=0$.
} \label{fig:peakPosition}
\end{figure}

\begin{figure}[t]
\includegraphics*[width=0.5\linewidth]{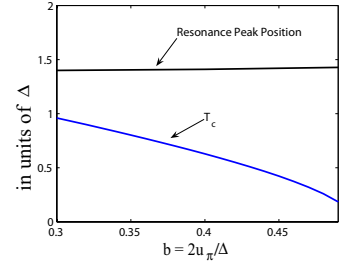}
\caption{Scattering by non-magnetic impurities. Evolution of $T_c$
and the resonance peak position (both in units of $\Delta$) as a
function of $b$.
} \label{fig:peakTc}
\end{figure}
In Fig.\ref{fig:peakPosition}(a)-(b) we show the calculated
Im$\chi(\Omega)$ for various values of $b$ and for two different
$u_0$. Im$\chi(\Omega)$ for $b=0$ is shown in panel (c) for
comparison.
We see that for  $b \sim 0.3-0.6$ used earlier to fit NMR,
penetration depth and uniform spin susceptibility data (see Refs.
\cite{chubukov,bib:vor}) the  resonance peak in Im$\chi({\bf
Q},\Omega)$ falls into the frequency range where the spin response
in a dirty $s^{\pm}$ SC is no longer a $\delta$-function, and its
width increases with increasing $b$. At the same time, the
position of the resonance  remains almost intact. Despite that
$\Delta_b$ and $T_c$ obviously decrease when $b$ increases. To
emphasize this point we plot in Fig.\ref{fig:peakTc}
$\Omega_{res}$ as a function of $b$ together with  $T_c (b)$
calculated from the conventional relation\cite{bib:vor}
\begin{equation}\label{Tc}
\ln\left( \frac{T_c}{T_{c_0}}\right)
=\psi\left(\frac{1}{2}\right)-\psi\left(\frac{1}{2}+ \frac{b \Delta}{2 \pi T_c}\right),
\end{equation}
where $\psi$ is the digamma-function and $T_{c0}$ is the transition temperature in the clean limit.
We clearly see that $T_c$ decreases with $b$ while $\Omega_{res}$ remains practically unchanged, such that the experimentally measured ratio
  $\Omega_{res}/T_c$ increases with the strength of intra-band impurity scattering.
This trend is consistent with the observed doping  dependence of
$\Omega_{res}/T_c$ in Ba$_{1-x}$K$_x$Fe$_2$As$_2$ above optimal
doping, where the decrease of $T_c$ with increasing $K$
concentration is believed to be at least partly  due to increased
impurity scattering~\cite{kogan}.

We next keep $u_0^p$ and $u_\pi^p$ non zero and compare the effects of non-magnetic and magnetic impurities.
%
\begin{figure}[t]$
\begin{array}{cc}
\includegraphics*[width=1.0\linewidth]{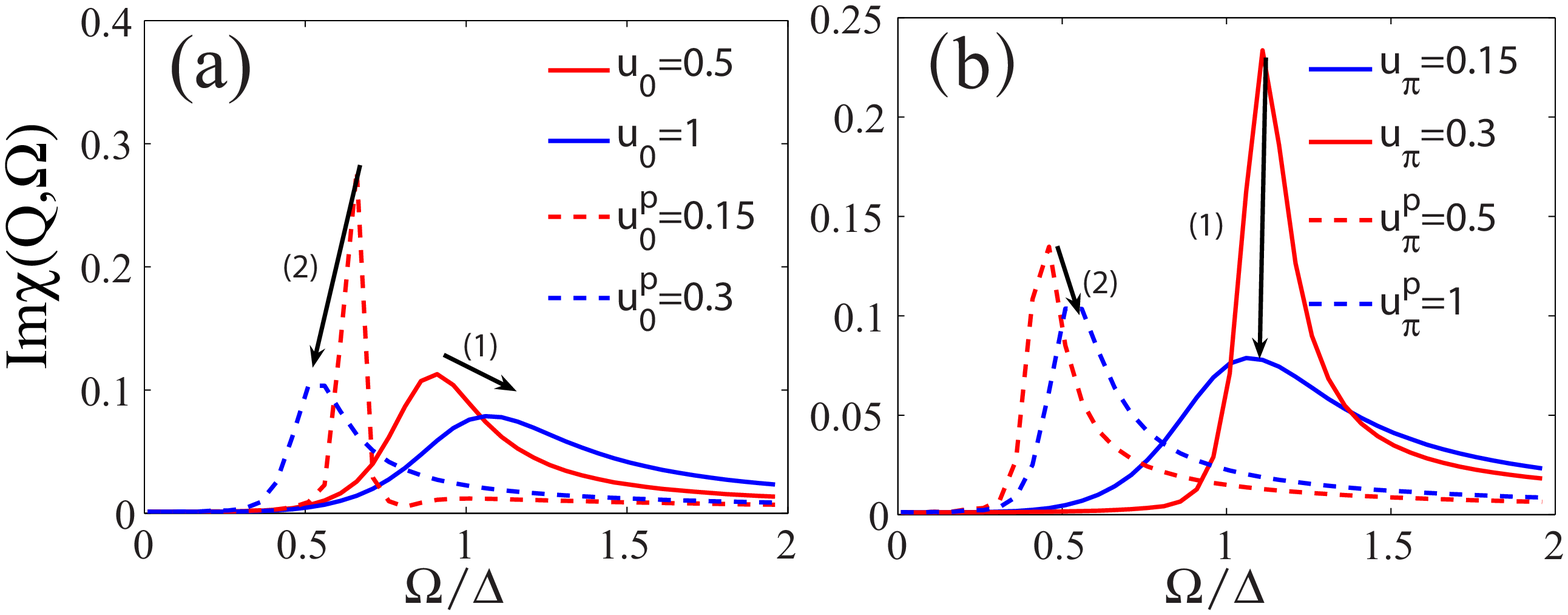}&
\end{array}$
\caption{ The effect of (a) intra-band and (b) inter-band impurity
scattering on the resonance peak in iron-based superconductors. In
particular, (a) shows the influence of the increasing intra-band
non-magnetic and magnetic scattering potential, while (b) displays
the corresponding effect of the inter-band impurity scattering. In
both cases, the solid and dashed curves refer to the non-magnetic
and magnetic impurity scattering, respectively. While studying the
effect of magnetic impurities, we kept the non-magnetic impurities
as constants and vice versa.} \label{fig7}
\end{figure}

In Fig.\ref{fig7}(a) we show how the resonance profile evolve once we increase
 small momentum impurity-scattering terms $u_0$ and $u_0^p$.  We see that in both cases
 the peak broadens but it is shifted in opposite directions -- upwards when $u_0$ increases and downwards when $u_0^p$ increases.
 Also, the broadening is much stronger under the change of $u_0^p$.
 Fig.\ref{fig7}(b) shows the effect on the peak of magnetic and non-magnetic
inter-band scattering $u_\pi$ and $u^p_\pi$.  We see that the increase of $u^\pi$ strongly broadens the peak and weakly shifts it downwards, while the increase of $u^p\pi$ has weaker effect on the width of the peak and shifts its position slightly upwards.
 This behavior is expected because, like we said,  in an $s^{\pm}$-wave superconductor
the interband non-magnetic impurity scattering $u_{\pi}$ behaves
 much like a magnetic intra-band scattering, $u_{0}^p$. Therefore
 $u_\pi$ and $u_0^p$ should give rise to qualitatively similar effects on the resonance
peak. This is indeed what we have found (compare arrow (2) in Fig. \ref{fig7}(a) and arrow (1) in
Fig.\ref{fig7}(b)). Likewise, the interband magnetic scattering,
$u_{\pi}^{p}$ and the intraband non-magnetic scattering $u_0$ should have similar effect on the resonance, and this also agrees with Fig.
\ref{fig7} (compare arrow (1) in Fig. \ref{fig7}(a) and arrow (2) in Fig.\ref{fig7}(b)).
One
has to bear in mind, however, that the effects of $u_\pi$ and $u^p_0$ and of $u_0$ and $u^p_{\pi}$ are similar but not equivalent because
 vertex corrections behave differently for  magnetic and non-magnetic
impurity scatterings (see Eq.(\ref{vertex_corr_chi_mag_imp})),
independent on whether this is intra- or inter-band scattering.
Still, the results in Fig. \ref{fig7} clearly show that
 pair-breaking $u_\pi$ and $u^p_0$ have the strongest effect on the resonance peak. Both impurity scatterings
  broaden the peak but only slightly affect its position such that $\Omega_{res}/T_c$ increases when the concentration of either
   non-magnetic or magnetic impurities increases.  This is the key result of this section.

\section{Conclusion}
\label{sec:6}

In this work we have shown that the intensity and position of the
resonance peak depends on several factors: doping, ellipticity of the electron pockets;
  gap anisotropy and the presence accidental nodes on the electron
pockets, and magnetic and non-magnetic
impurity scattering.

We find that with no ellipticity, the resonance
starts at the commensurate momentum below $2\Delta$ and moves
outward  in energy with doping. At some doping, the minimum in the dispersion of the resonance jumps to an incommensurate momentum,  and at even larger doping the intensity of the resonance becomes the largest at the incommensurate momentum.
 With finite ellipticity, the intensity becomes the largest at the incommensurate momenta at smaller dopings, and for large enough ellipticity
 the resonance becomes incommensurate already at zero doping. Besides, at finite ellipticity, the resonance peak gets quite broad  The inclusion of the angular variation of the $s^{\pm}-$
 leads to further broadening of the resonance peak, particularly when the gap has accidental nodes
on the electron pockets.  The broadening of the resonance may actually be even stronger than in $d-$wave cuprate superconductors.

We further analyzed the sensitivity of the resonance to non-magnetic as well as
magnetic impurities. We obtain that both non-magnetic and magnetic
impurities have pair-breaking components which broaden the resonance but do not shift much its position compared to the clean case.
The reduction of  $T_c$ by the same impurity scattering is much stronger, such that
the ratio $\Omega_{res}$/T$_c$ increases with the strength of impurity scattering.

The transformation of the minimum in the dispersion of the resonance and the maximum of its intensity  from the commensurate to the incommensurate momentum upon doping is consistent with recent study of the doping evolution of the resonance in
Ba$_{1-x}$K$_x$Fe$_2$As$_2$~\cite{osborn}. The increase of $\Omega_{res}/T_c$ with the increase of impurity scattering is consistent with the
 observed increase of this ratio with doping in overdoped systems, where the decrease of $T_c$ with doping can at least partly be attributed to the effect of impurities.

We thank R. Fernandes, M. Korshunov, M. Norman, R. Osborn, M. Vavilov, A. Vorontsov,  for useful conversations.
A.V.C. acknowledges the support from NSF-DMR 0906953  and is thankful to MPIPKS in Dresden for hospitality during the  completion of the manuscript. JK acknowledges
support from a Ph.D. scholarship from the Studienstiftung
des deutschen Volkes and the IMPRS Dynamical Processes
in Atoms, Molecules and Solids.
IE acknowledges the DAAD
PPP Grant No.50750339.

\section{Appendix}

\section{s$^{\pm}$ versus s$^{++}$-wave gap}
\label{sec:5}

In this Appendix, we briefly address the issue of the interplay
between spin response in superconductors with $s^{\pm}$ and
$s^{++}$ gaps in a situation when the original and the shadow FSs
do not cross, which is the case when  both hole and electron  FSs
are near-circular and the doping is finite, such that one FS is
larger than the other.   The case when the original FSs cross is
quite similar to the cuprates, and the absence of spin resonance
for $s^{++}$ gap in this situation has been discussed
earlier~\cite{second_res}.

We show that there is no resonant enhancement of the Im$\chi$ for
the $s^{++}$-wave case even when the original and the shadow FSs
do not cross. To do so we look into the case with no impurity
scattering and no ellipticity and compute \beq \label{1}
\chi^{\alpha\beta}_0(Q,i\Omega) = - N_0 T\sum_n \int d\varepsilon
G^{\alpha}(i\omega_n,\varepsilon)G^{\beta}(i\omega_n+\Omega,\varepsilon+\delta\mu)
\eeq

where $\delta\mu$ is proportional to the mismatch in FS radii due
to doping. And $\alpha$ and $\beta$ refer to the hole and electron
FSs respectively. In the absence of impurities we return to the
case where ${\bar\Delta}_\omega$ is frequency independent and
simply equal to $\Delta$. In the normal state we get \beq\label{2}
Im\chi_0 (Q,\Omega) =
N_0\frac{\pi^2}{2}\theta(\Omega-|\delta\mu|)\eeq In the SC state
(after appropriately adding the F*F term), we get \beq\label{3}
Im\chi_0 (Q,\Omega) = -N_0
\frac{\pi^2}{4}\sum_{x}\frac{\Delta(\Delta+\Delta_Q)+(\omega+x)(\Omega-\delta\mu)}{|\Omega
x-\delta\mu~\omega|} \eeq
where  $\Delta$ and $\Delta_Q$ are the gaps on the hole
and electron FSs, and the two values of $x$ which are summed over
are given by
\beq\label{4}
x=-\frac{\delta\mu}{2}\pm\frac{\Omega}{2}\sqrt{\frac{\left(\frac{\Omega}{2}\right)^2-\left(\frac{\delta\mu}{2}\right)^2-\Delta^2}{\left(\frac{\Omega}{2}\right)^2-\left(\frac{\delta\mu}{2}\right)^2}}
\eeq

For zero doping ($\delta\mu=0$), for $s^{\pm}$ SC,
$\Delta=-\Delta_Q$ and we have \beq\label{5} Im\chi^{s^{\pm}}_0 =
N_0
\frac{\pi^2}{2}\frac{1}{\sqrt{1-\left(\frac{2\Delta}{\Omega}\right)^2}}\eeq
\begin{figure}[!t]
\includegraphics[width=1.0\linewidth]{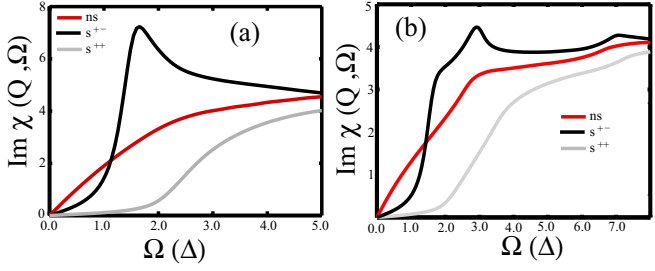}
\caption{(color online) Calculated frequency dependence of the
imaginary part of the total physical susceptibility for
(a)$\epsilon=0.3$ and (b)$\epsilon=0$ for the $s^{++}$-wave
superconductor. We set $x=0.01$ ($\delta\mu=-0.02$). We clearly
see that there is no resonant enhancement of the full spin
susceptibility for $s^{++}$ gap.  To soften $\delta-$function in
the $s^{\pm}$ case we averaged Im $\chi ({\bf Q}, \Omega)$ around
${\bf Q}$ with $\delta {\bf q} = \pm 0.1 {\bf Q}$ }
\label{fig6new}
\end{figure}

This implies that Re$\chi_0$ has a square-root singularity at
$\Omega=2\Delta$ and hence the full Im$\chi^{s\pm} ({\bf Q},
\Omega) $ has a pole (resonance)
 at some $\Omega < 2\Delta$.

For $s^{++}$ case $\Delta=\Delta_Q$, and we get \beq\label{6}
Im\chi^{s++}_0 =N_0
\frac{\pi^2}{2}\sqrt{1-\left(\frac{2\Delta}{\Omega}\right)^2}\eeq
As expected,  there is no jump in $Im\chi^{s++}_0 ({\bf Q},
\Omega)$  at $2\Delta$ implying no singularity in $Re\chi_0$ and
hence no resonance in Im $\chi^{s++} ({\bf Q}, \Omega)$ below
$\Omega=2\Delta$.

Consider next the case when $\delta \mu$ is non-zero. Now
Im$\chi_0(Q,\Omega)$ has a singularity at
$\Omega=\sqrt{\delta\mu^2+4\Delta^2}$. Expanding around this
$\Omega$ by introducing a small $z$ such that
$\left(\frac{\Omega}{2}\right)^2=\left(\frac{\delta\mu}{2}\right)^2+\Delta^2+z^2$,
we find using Eq. \ref{3}, that

\beq\label{7} Im\chi^{s^{\pm}}_0 (z) =N_0
\frac{\pi^2}{2}\frac{\Delta}{z}\eeq

and

\beq\label{8} Im\chi^{s++}_0 (z) = N_0\frac{\pi^2}{2}z\eeq

So the main result is the same as with no doping  -- Im
$\chi^{s\pm}_0 (z)$ is singular at $z=0$,  the singularity in Im
$\chi_0 ^{s\pm} (0)$ gives rise to the singularity in Re
$\chi^{s++}_0 (0)$ and to resonance at a smaller frequency. On the
other hand, Im $\chi^{s++}_0 (z)$ is non-analytic at $z=0$, but
not singular, hence Re $\chi^{s++}_0 (0)$ is also not singular,
and Im $\chi^{s++}_0 (z)$  remains non-singular below the bottom
of the particle-hole continuum. We further confirm these
statements by performing the numerical calculations for our
four-band model. We show the results in Fig.\ref{fig6new}.


\begin{thebibliography}{99}
\bibitem{comm} We remind that FeTe$_{0.6}$Se$_{0.4}$~
 is very different  from a parent compound  FeTe
and shows a metallic behavior in the normal state and a spin
response similar to that in Fe-pnictides.
\end{thebibliography}
\end{document}